\documentclass[journal=langd5,manuscript=article]{achemso}
\usepackage{graphics,graphicx,mathrsfs,color,amsmath}
\usepackage{chemformula} 
\usepackage[T1]{fontenc} 

\newcommand{\ud}{\mathrm{d}}
\newcommand{\change}[1]{\textcolor{black}{#1}}
\newcommand{\zj}[1]{\textcolor{black}{#1}}

\author{Tian Yu}
\affiliation{Center of Soft Matter Physics and Its Applications, Beihang University, 37 Xueyuan Road, Beijing 100191, China}
\alsoaffiliation{School of Physics and Nuclear Energy Engineering, Beihang University, 37 Xueyuan Road, Beijing 100191, China}

\author{Ying Jiang}
\email{yjiang@buaa.edu.cn}
\affiliation{Center of Soft Matter Physics and Its Applications, Beihang University, 37 Xueyuan Road, Beijing 100191, China}
\alsoaffiliation{School of Chemistry, Key Laboratory of Bio-Inspired Smart Interfacial Science and Technology of Ministry of Education, Beihang University, 37 Xueyuan Road, Beijing 100191, China}
\alsoaffiliation{Beijing Advanced Innovation Center for Biomedical Engineering, 37 Xueyuan Road, Beihang University, Beijing 100191, China}

\author{Jiajia Zhou}
\email{jjzhou@buaa.edu.cn}
\affiliation{Center of Soft Matter Physics and Its Applications, Beihang University, 37 Xueyuan Road, Beijing 100191, China}
\alsoaffiliation{School of Chemistry, Key Laboratory of Bio-Inspired Smart Interfacial Science and Technology of Ministry of Education, Beihang University, 37 Xueyuan Road, Beijing 100191, China}
\alsoaffiliation{Beijing Advanced Innovation Center for Biomedical Engineering, 37 Xueyuan Road, Beihang University, Beijing 100191, China}

\author{Masao Doi}
\affiliation{Center of Soft Matter Physics and Its Applications, Beihang University, 37 Xueyuan Road, Beijing 100191, China}

\title{Dynamics of Taylor Rising}

\begin{document}

\newpage
\begin{abstract}
We study the dynamics of liquid climbing in a narrow and tilting corner, inspired by recent work on liquid transportation on the peristome surface of Nepenthes alata. 
Considering the balance of gravity, interfacial tension and viscous force, we derive a partial differential equation for the meniscus profile, and numerically study the behavior of the solution for various tilting angle $\beta$. 
We show that the liquid height $h(t)$ at time $t$ satisfy the same scaling law found for vertical corner, i.e., $h(t) \propto t^{1/3}$ for large $t$, but the coefficient depends on the tilting angle $\beta$. 
The coefficient can be calculated approximately by Onsager principle, and the result agrees well with that obtained by numerical calculation. 
Our model can be applied for a weakly curved corner and may provide guidance to the design of biomimetic surfaces for liquid transportation.

\end{abstract}

\section{Introduction}
It is well-known that when a capillary is inserted to a bulk liquid, the liquid rises in the tube \cite{dBQ}. 
A similar phenomenon is observed when a wedge shape consisting of two intersecting plates is in contact with the bulk liquid. 
A liquid finger quickly forms and climbs along the corner.
The earliest study of the wedge system dates back to the 18th century.
In 1712, Brook Taylor conducted experiments on the capillary rising at a small-angle wedge formed by two nearly parallel plates. 
He found that the equilibrium shape of the meniscus is a hyperbola \cite{Taylor1710}. 
This observation was confirmed and quantified by Francis Hauksbee \cite{Hauksbee1710}. 
Two centuries later, Concus and Finn proposed the condition for the liquid to ascend \cite{Concus1969}. 
They showed that the liquid will wet the corner and rise along the edge only when the contact angle $\theta$ of the liquid and the open angle $\alpha$ of the plates satisfy
\begin{equation}
  \theta + \frac{\alpha}{2} < \frac{\pi}{2}.
\end{equation}
Other reports on the equilibrium meniscus can be found in Refs. \cite{Langbein1990, Finn1999, Finn2002a}.

Compared with the equilibrium theory of meniscus, theory on the dynamics of meniscus \change{started} much later.
The time evolution of the meniscus is determined by two competing effects.
On the one hand, the liquid tends to wet the plate surfaces to minimize the interfacial energy, while on the other hand, the liquid has to overcome the gravity to ascend, and the liquid flow is also slowed down by the viscosity.
When the effect of gravity can be neglected, the propagation of the meniscus front obeys the classical Lucas-Washburn's $t^{1/2}$ scaling \cite{Lucas1918, Washburn1921}.
This has been observed in capillary rising in a microgravity environment \cite{Weislogel2001} or in the imbibition taking place horizontally \cite{Ayyaswamy1974, Dong1995}.
When gravity is considered, the meniscus rises with a different $t^{1/3}$ scaling.
This result was first derived by Tang and Tang \cite{Tang1994}.
Higuera et al. developed a more complete theory for the case of two vertical plates forming a small angle \cite{Higuera2008}.
They derived a partial differential equation for the time evolution of the meniscus shape at the late stage and derived the $t^{1/3}$ scaling law from this equation.
Ponomarenko et al. conducted experiments of capillary rising in the corners of different geometries and demonstrated the scaling $t^{1/3}$ is universal.
Recently, a study on the peristome surface of Nepenthes alata showed that the plant has taken full advantage of the corner geometry to realize the directional control of the liquid flow \cite{ChenHuawei2016, ChenHuawei2017}.
The microstructures on the peristome surface resembles the geometry of intersecting plates, but with the spine curved and tilted. 

In this paper, we propose a simple model on the capillary flow at a narrow corner. 
We start with a general system where two intersecting plates are inserted to a liquid bath with a titling angle $\beta$.
Using Onsager principle \cite{DoiSoft}, we derive a partial differential equation that describes the time evolution of liquid climbing along the corner.
We obtain both numerical solutions and approximate analytical solutions to the partial differential equation. 
Our results show that the length of the meniscus follows a time-scaling of $t^{1/3}$, which is consistent with previous studies \cite{Tang1994, Higuera2008, Ponomarenko2011}.
To mimic the microstructure on Nepenthes alata, we extend this model to the case that the intersecting line of the two plates is curved. 
We expect this work could provide guidance to the design of capillary flows in systems with complex geometry.

\section{Theoretical Model}

\begin{figure}[htp]
\includegraphics[width=1.0\textwidth]{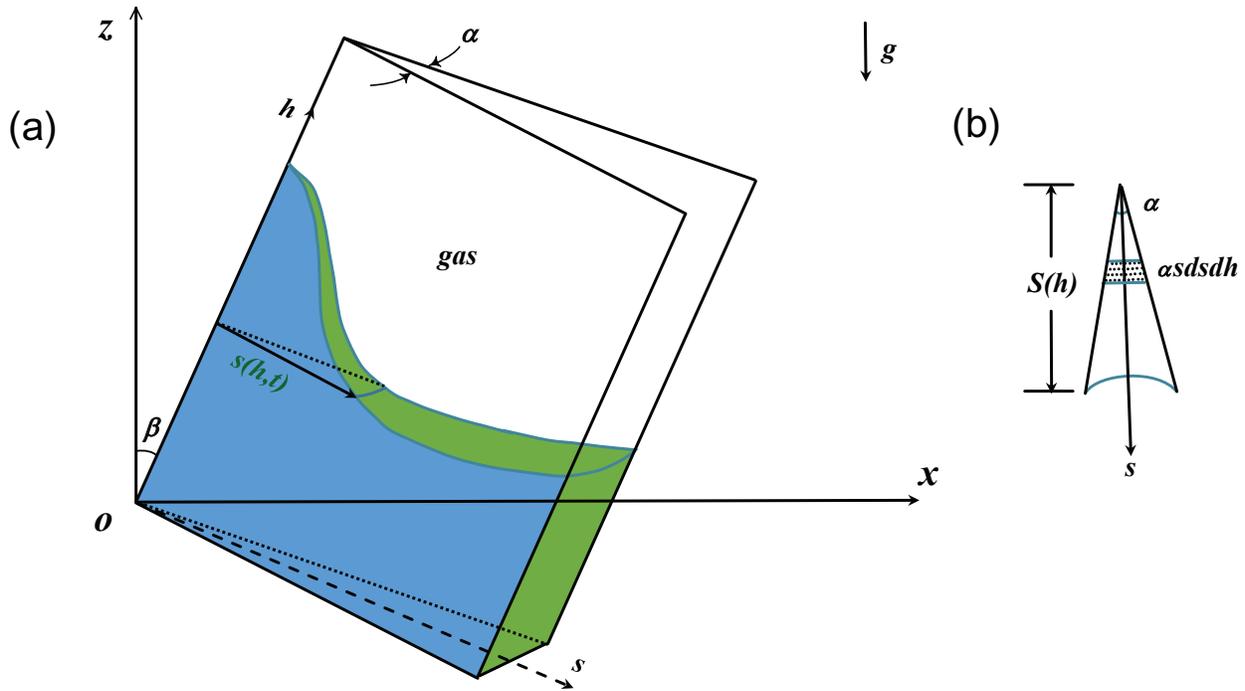}
\caption{(a) Schematic picture of capillary flow of liquid confined by a tilting corner. $\alpha$: intersection angle of the two plates; $\beta$: inclined angle of the spine; $h$: the coordinate along the spine; $s$: the coordinate representing the distance from the spine; $g$: gravitational constant.
(b) A small volume element of liquid we used for calculation. See text for details.}
\label{fig:1}
\end{figure}

We study a model system shown schematically in Fig.~\ref{fig:1}(a). 
Two plates are intersected with a small angle $\alpha$, and the intersecting line is denoted as $h$-axis.
The axis that is perpendicular to the $h$-axis and bisects the $\alpha$ angle is denoted as $s$-axis.  
The two plates are inserted into a semi-infinite liquid pool at $z<0$ with a tilting angle $\beta$. 
We will use two Cartesian coordinates: one is $x$-$z$, with $z$-axis pointing upward, in the opposite direction of gravity $g$; the other one is $s$-$h$ which is  associated to the plates' frame.
The density of the liquid, the viscosity, and the surface tension are denoted by $\rho$,  $\eta$, and $\gamma$, respectively. 

The equilibrium contact angle of the liquid to the plates is $\theta$. 
If the condition $\theta + \alpha/2 < \pi/2$ is satisfied, the liquid wets the plates and rises along the intersecting corner \cite{Concus1969}. 
The meniscus profile can be expressed by a function $h=H(s)$, or by its inverse function $s=S(h)$.

\section{Results and Discussions}

\subsection{Equilibrium Shape of Meniscus}

The equilibrium profile is determined by minimizing the total energy of the liquid. 
Consider a small volume element of liquid shown in Fig.~\ref{fig:1}(b). 
The mass of the liquid is $\rho s \alpha \ud s \ud h$ and the height is $ h\cos\beta - s \sin\beta$. 
\zj{Setting the $z=0$ plane as the reference state, the gravitational energy of the cross-section is then given by}
\begin{equation}
  \int_0^{S(h)} \ud s \rho g s \alpha ( h \cos\beta - s \sin\beta) \ud h
  = \rho g \alpha \left( \frac{1}{2} S^2 h \cos\beta - \frac{1}{3} S^3 \sin\beta \right) \ud h .
\end{equation}
The total gravitational energy is 
\begin{equation}
  \label{eq:F_grav}
  F_{\rm gravity}= \int \ud h \, \rho g \alpha 
  \left( \frac{1}{2} S^2 h \cos\beta - \frac{1}{3} S^3 \sin\beta \right) .
\end{equation}

\zj{Let $\gamma_{SL}$ and $\gamma_{SV}$ be the interfacial tension between  solid-liquid and solid-vapor interfaces. 
For a surface area $2S\ud h$, the difference in the interfacial energy  between the wetted state and the state exposed to the vapor is $2 S \ud h(\gamma_{SL}-\gamma_{SV})$. 
Using Young-Dupre relation $\gamma_{SV} = \gamma_{SL} + \gamma \cos\theta$, the interfacial energy can be written as $2 S \ud h (-\gamma\cos\theta)$.}
Here we have neglected the contribution from the liquid/air interface because the angle $\alpha$ is small.
The total interfacial free energy is then
\begin{equation}
  \label{eq:F_int}
  F_{\rm interface} = - \int \ud h \, 2 S \gamma \cos\theta .
\end{equation}

The total energy is given by the summation of the gravitational energy (\ref{eq:F_grav}) and the interfacial energy (\ref{eq:F_int})
\begin{equation}
  \label{eq:F_tot}
  F = \int \ud h \bigg[ \rho g \alpha \left( \frac{1}{2} S^2 h \cos\beta 
    - \frac{1}{3} S^3 \sin\beta \right) - 2 S \gamma \cos\theta \bigg] .
\end{equation}
The meniscus profile at equilibrium can be determined by minimizing the total energy with respect to the meniscus profile $S(h)$.
This leads to 
\begin{equation}
  \rho g \alpha ( S h \cos\beta - S^2 \sin\beta ) - 2 \gamma \cos\theta = 0.
\end{equation}

It is convenient to express the profile using $H(s)$, the inverse function of $S(h)$. 
The equilibrium shape is given by
\begin{equation}
  \label{eq:eqH}
  H(s) = s \tan\beta+\frac {2\gamma\cos\theta}{\rho g\alpha \, s\cos\beta}.
\end{equation}

\zj{We choose a reference length $H_c=(2\gamma\cos\theta / \rho g\alpha)^{1/2}$ \cite{Higuera2008}.
This reference length plays the role of capillary length in our system.
For the specific case of $\beta=0$, suppose that a small volume element of liquid with a height $h$ and a length $s$, the gravitational energy dominates when $h s \gg H_c^2$, while the interfacial energy is more important in the opposite limit of $ h s \ll H_c^2$.} 
We make the equations dimensionless by scaling all \change{lengths} with $H_c$.
The equilibrium profile (\ref{eq:eqH}) then takes the form  
\begin{equation}
  \label{eq:eqH2}
  \tilde{H} (\tilde{s}) = \tilde{s}\tan\beta+\frac {1}{\cos\beta}\frac {1}{\tilde{s}}.      
\end{equation}
All variables with a tilde are dimensionless quantities.
For the special case of $\beta=0$, the plates are vertically inserted into the liquid, Eq.~(\ref{eq:eqH2}) becomes a hyperbola
\begin{equation}
  \beta=0: \quad \tilde{H}(\tilde{s}) =\frac {1}{\tilde{s}}.
\end{equation}
This is a well-known result dated back to Taylor and Hauksbee \cite{Taylor1710, Hauksbee1710}. 

For an arbitrary tilting \change{angle} $\beta$,  we can rewrite Eq.~(\ref{eq:eqH2}) in the $x$-$z$ coordinates
\begin{equation}
  \tilde{x}=\tilde{z}\tan\beta+\frac {1}{\cos\beta}\frac {1}{\tilde{z}}\, .      
\end{equation}
This equation can be solved for $\tilde{z}$, which gives the meniscus height as a function of $\tilde{x}$. 
The solution has two branches
\begin{eqnarray}
  \label{eq:z1}
  \tilde{z}_1 &=& \frac{ \tilde{x} + \sqrt{ \tilde{x}^2 
                  - 4 \tan \beta / \cos\beta }}{2\tan\beta} , \\
  \label{eq:z2}
  \tilde{z}_2 &=& \frac{ \tilde{x} - \sqrt{ \tilde{x}^2
                  - 4 \tan \beta / \cos\beta }}{2\tan\beta} .
\end{eqnarray}
When $\tilde{x}$ goes to infinity, $\tilde{z}_1$ approaches to the spine, $\tilde{z}=\tilde{x}\cot\beta$, while $\tilde{z}_2$ approaches to the surface of the liquid bath $\tilde{z}=\frac {1}{\cos\beta}\frac {1}{\tilde{x}}$.
These two branches meet at point 
\begin{equation}
  \tilde{x}_0 = 2 \sqrt{ \frac{ \tan\beta }{ \cos\beta }}, \quad 
  \tilde{z}_0 = \sqrt{ \frac{1}{ \sin\beta } }.
\end{equation}
These profiles will be shown later (see Fig. \ref{fig:2}).

\subsection{Time Evolution Equation for Meniscus}

In this section, we consider the time evolution of the meniscus. 
Higuera et al. have shown that at the initial stage of the meniscus rise, the effect of gravity is negligible \cite{Higuera2008}. 
In the region far away from the corner, the meniscus quickly approaches to the equilibrium profile.
Therefore we assume that the liquid in the region of $z<z_0$ is in equilibrium (given by the Eq. (\ref{eq:z2})), and focus our attention to the dynamics in the region $z>z_0$.


The time evolution of the meniscus can be derived by Onsager principle \cite{DoiSoft, XuXianmin2016, DiYana2016, DiYana2018, YuTian2018}.
The time evolution of the meniscus can be determined by the minimum of the Rayleighian defined by
\begin{equation}
  \mathscr{R} = \dot{F} + \Phi ,     
\end{equation}
where $\dot{F}$ represents the time derivative of the free energy of the liquid and $\Phi$ is the dissipation function, equal to half of the energy dissipated per unit time \cite{DoiSoft}.

The free energy can be obtained through Eq.~(\ref{eq:F_tot})
\begin{equation}
  F = \int_{h_0}^{h_m(t)} \ud h \Big[\rho g \alpha 
  \left( \frac{1}{2} S^2 h \cos\beta -\frac{1}{3} S^3 \sin\beta \right)
  -2\gamma S \cos\theta \Big]\, ,     
\end{equation}
where $S(h,t)$ is the time-dependent profile of the meniscus, $h_m(t)$ is the position of the meniscus front along the spine at time $t$, i.e., $S(h=h_m,t)=0$, and $h_{0}$ is the $h$ coordinate of the intersection point $(x_0,z_0)$. 
The time derivative of the free energy is
\begin{equation}
  \label{eq:Fdot}
  \dot{F} = \int_{h_0}^{h_m(t)} \ud h \Big[ \rho g \alpha
  \left( S h \cos\beta - S^2 \sin\beta \right)
  -2 \gamma \cos\theta \Big] \dot{S} .
\end{equation}

Let $Q(h,t)$ be the volume flux of liquid flowing across the plane at $h$. 
The volume conservation equation for the liquid is written as
\begin{equation}
  \label{eq:vol_cons}
  \alpha S \dot{S} = -\frac {\partial Q}{\partial h}\, .     
\end{equation}
Substituting Eq.~(\ref{eq:vol_cons}) into Eq.~(\ref{eq:Fdot}), we get
\begin{eqnarray}
  \dot{F} &=& \int_{h_{0}}^{h_m(t)} \ud h 
     \Big[ - \rho g (h\cos\beta-S\sin\beta )+\frac {2\gamma\cos\theta}{\alpha S}\Big]
              \frac {\partial Q}{\partial h} \nonumber \\
  &=& \int_{h_{0}}^{h_m(t)} \ud h 
     \Big[ \rho g (\cos\beta- \sin\beta \frac{\partial S}{\partial h} )
      +\frac {2\gamma\cos\theta}{\alpha S^2} \frac{\partial S}{\partial h} \Big] Q,
\end{eqnarray}
where integration-by-part has been used in the last step.

The dissipation function $\Phi$ can be calculated by the lubrication approximation \cite{Batchelor}. 
Let $v_h(h,s,t)$ and $v_s(h,s,t)$ be the depth-averaged velocity of fluid in $h$- and $s$-direction at position $(h,s)$ and time $t$.  
In the lubrication approximation, the fluid velocity $\boldsymbol{v}=(v_h,v_s)$ is proportional to the pressure gradient $\nabla p$ in the fluid, 
\begin{equation}
  \boldsymbol{v} = - \frac{1}{\xi} \nabla p ,
  \label{eqn:19a}
\end{equation}
where $\xi$ is the \change{friction} constant for the fluid motion at point $(h,s)$.   
$\xi$ is expressed in terms of the fluid viscosity $\eta$ and the thickness $e$ of the fluid
\begin{equation}
  \xi = \frac{\eta}{12 e^2} .
  \label{eqn:19b}
\end{equation}
Since the fluid is moving in a narrow region along the spine, we may ignore $v_s$ and assume that $p$ depends on $h$ only. 
\zj{In Ref.~\cite{Higuera2008}, Higuera et al. kept both $v_h$ and $v_s$ and derived evolution equations which are also valid for the initial times. 
Since we are interested in the long-time behavior, we shall make this simple one-dimension approximation and keep only $v_h$ component.} 
Equations (\ref{eqn:19a}) and  (\ref{eqn:19b}) then indicate that $v_h$ is written as 
\begin{equation}
  v_h(h,s,t)=C(h,t) s^2, 
  \label{eqn:19c}
\end{equation}
where we have used $e=\alpha s$. 
The volume flux $Q$ can be written as
\begin{equation}
  \label{eq:Q}
  Q=\int_0^{S} \ud s \alpha s v_h .     
\end{equation}
Form Eqs. (\ref{eqn:19c}) and (\ref{eq:Q}), $v_h$ can be written as
\begin{equation}
  v_h=\frac {4Q}{\alpha S^4} s^2 .
\end{equation}
The energy dissipation function $\Phi$ is then given by
\begin{equation}
  \label{eq:Phi}
  \Phi=\frac {1}{2} \int_{h_{0}}^{h_m(t)} \ud h 
  \int_0^{S} \ud s \frac {12\eta}{\alpha s} v_h^2
  =\frac {1}{2}\int_{h_0}^{h_m(t)} \ud h \frac{48\eta}{\alpha^3 S^4} Q^2 .         
\end{equation}

The Rayleighian of the system is given by the summation of the time derivative of the free energy (\ref{eq:Fdot}) and the dissipation function (\ref{eq:Phi})
\begin{equation}
  \mathscr{R} = \dot{F} + \Phi 
  = \int_{h_{0}}^{h_m(t)} \ud h 
     \left\{ \Big[ \rho g (\cos\beta- \sin\beta  \frac{\partial S}{\partial h})
      +\frac {2\gamma\cos\theta}{\alpha S^2} \frac{\partial S}{\partial h} \Big] Q
      + \frac{1}{2} \frac{48 \eta}{\alpha^3 S^4} Q^2 \right\}
\end{equation}
The time evolution equation is then given by $\delta \mathscr{R}/\delta Q=0$,
\begin{equation}
  \label{eq:Q2}
  Q = \frac {\alpha^3 S^4}{48\eta} \left[-\rho g\cos\beta
    +\left(\rho g\sin\beta-\frac {2\gamma\cos\theta}{\alpha S^2}\right) 
    \frac {\partial S}{\partial h}\right] .   
\end{equation}
Substituting the conservation equation (\ref{eq:vol_cons}) into Eq.~(\ref{eq:Q2}), we arrive at the following time evolution equation for the meniscus profile $S(h)$
\begin{eqnarray}
  \frac {\partial S}{\partial t} &=& 
    \frac{\alpha^2 S^2}{12\eta} \rho g \cos\beta \frac{\partial S}{\partial h}
    + \left[ \frac{\alpha \gamma \cos\theta}{12\eta} 
            -\frac{\alpha^2 S^2}{12 \eta} \rho g \sin\beta \right] 
       \left( \frac{\partial S}{\partial h} \right)^2 \nonumber \\
  \label{eq:evolution0}
  & +& \left[ \frac{\alpha \gamma \cos\theta S}{24\eta} 
            -\frac{\alpha^2 S^3}{48\eta} \rho g \sin\beta \right]
       \frac{\partial^2 S}{\partial h^2} .
\end{eqnarray}

\zj{We can make the equation dimensionless by scaling $h$ and $S$ with $H_c = (2\gamma\cos\theta / \alpha \rho g)^{1/2}$ and the time $t$ with $t_c= 12 \eta / \alpha^2 \rho g H_c$ for corresponding dimensionless form $\tilde{h}$, $\tilde{S}$, $\tilde{t}$, respectively. 
We shall see late that when $t=t_c$, the meniscus climbs up to a height on the order of $H_c$.
The evolution equation then becomes}
\begin{equation}
  \label{eq:evolution}
  \frac {\partial \tilde{S}}{\partial \tilde{t}} = 
  \tilde{S}^2 \cos\beta \frac{\partial\tilde{S}}{\partial \tilde{h}}
  + \left( \frac{1}{2} - \tilde{S}^2\sin\beta \right) 
  \left(\frac {\partial \tilde{S}}{\partial \tilde{h}} \right)^2 
  + \left( \frac{1}{4} \tilde{S} -  \frac{1}{4} \tilde{S}^3 \sin\beta \right) 
  \frac{\partial^2 \tilde{S}}{\partial \tilde{h}^2} .
\end{equation}

The above equation can be solved numerically with suitable initial condition.
Different initial conditions converge to almost the same profiles after a short period of time, which lead to the same long-time dynamics.
We choose a straight line as the initial condition for a numerical calculation of Eq.~(\ref{eq:evolution}). 
(The effect of the initial condition is discussed in more detail in the Appendix.)
Figure~\ref{fig:2} shows the time evolution of the meniscus profiles for different tilting angle $\beta$. 
For $\beta=0$ (the case of vertical spine), the results agree with that of Higuera et al. \cite{Higuera2008} \change{at long times.}

\begin{figure}[htp]
  \includegraphics[width=1.0\textwidth]{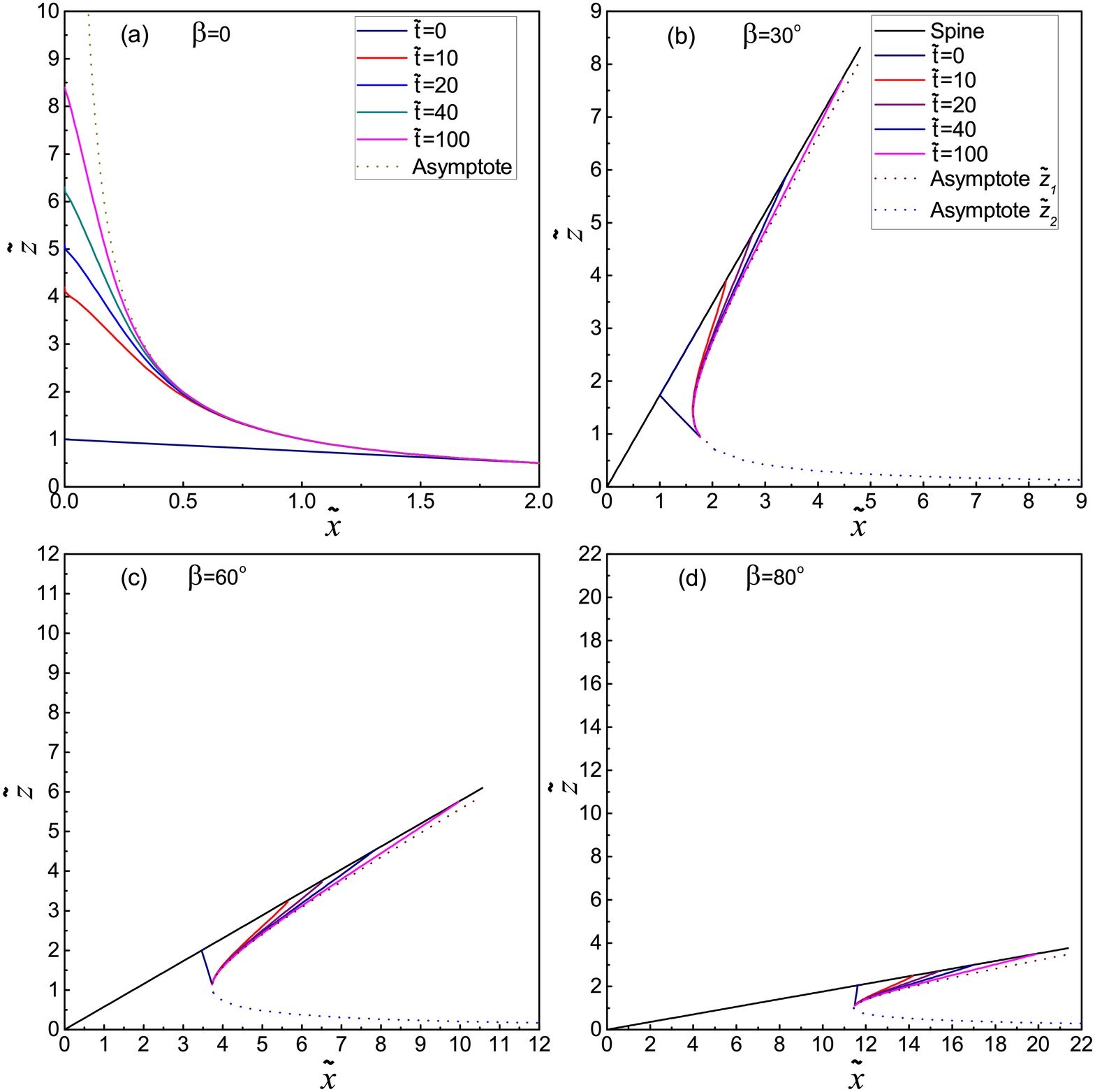}
  \caption{Profiles of time evolution of meniscus for different tilting angle: for 
(a) $\beta=0^{\circ}$, (b) $\beta=30^{\circ}$, (c) $\beta=60^{\circ}$, (d) $\beta=80^{\circ}$. 
The dot lines denote corresponding equilibrium states (\ref{eq:z1}) and (\ref{eq:z2}).}
  \label{fig:2}
\end{figure}

\subsection{Approximate Solution}
\label{sec:CPBS}

In this section we propose an analytical approach which uses Onsager principle as an approximation method\cite{Doi2015}.              
Equation (\ref{eq:eqH}) shows that the equilibrium meniscus profile can be well approximated by $H=H_c^2/ (s\cos\beta)$. 
Therefore we assume that the meniscus keeps this functional form in the transient state, and write $H(s,t)$ as
\begin{equation}
  \label{eq:CPBS}
  H(s,t) = \frac {H_c^2}{[s+a(t)]\cos\beta} ,
\end{equation}
where $a(t)$ is a certain function of time, which we shall determine in the following.

Since the function $H(s,t)$ satisfies the condition $H(0,t)=h_m(t)$, $a(t)$ can be expressed by $h_m(t)$ as 
\begin{equation}
  \label{eq:at}
  a(t)=\frac {H_c^2}{h_m(t)\cos\beta} .
\end{equation}
Substituting Eq.~(\ref{eq:at}) into Eq.~(\ref{eq:CPBS}), and again we convert to $S(h)$ expression instead of $H(s)$ for the meniscus profile, 
\begin{equation}
  \label{eq:S_CPBS}
  S(h,t) = \frac {H_c^2}{\cos\beta} \left( \frac {1}{h}-\frac {1}{h_m(t)} \right). 
\end{equation}

The free energy can be calculated from Eqs.~(\ref{eq:F_tot}) and (\ref{eq:S_CPBS}),
\begin{eqnarray}
  F &=& \int_{h_0}^{h_m(t)} \ud h \bigg[ \rho g \alpha \left( \frac{1}{2} S^2 h \cos\beta 
    - \frac{1}{3} S^3 \sin\beta \right) - 2 S \gamma \cos\theta \bigg] \nonumber \\
  &=& \frac{1}{2} \alpha \rho g \frac{H_c^4}{\cos\beta} 
      \left[ \textcolor{black}{\ln \frac{h_m}{h_0}} - \frac{2}{h_m} (h_m-h_0)
      + \frac{1}{2 h_m^2} (h_m^2 - h_0^2) \right] \nonumber \\
  && - \frac{1}{3} \alpha \rho g \frac{H_c^6 \sin\beta}{\cos^3\beta} 
      \left[ -\frac{1}{2} (\frac{1}{h_m^2} - \frac{1}{h_0^2})
        + \frac{3}{h_m} (\frac{1}{h_m} - \frac{1}{h_0}) 
        + \frac{3}{h_m^2} \ln\frac{h_m}{h_0} 
        - \frac{1}{h_m^3} (h_m - h_0) \right] \nonumber \\
  && - 2\gamma\cos\theta \frac{H_c^2}{\cos\beta}  
      \left[ \textcolor{black}{\ln \frac{h_m}{h_0}} - \frac{1}{h_m} (h_m-h_0) \right].
\end{eqnarray}

Since we are interested in the long-time dynamics, $h_m(t) \gg h_0$, the dominate terms in the free energy are $\mathcal{O}( \ln \frac{h_m}{h_0} )$ and $\mathcal{O}( \frac{1}{h_0^2} )$. 
We keep only these terms in the following calculation of the time derivative.
The time derivative of the free energy is
\begin{equation}
  \label{eq:dotF_CPBS}
  \dot{F} = \left( \frac{1}{2} \alpha \rho g \frac{H_c^4}{\cos\beta} 
    - 2\gamma\cos\theta \frac{H_c^2}{\cos\beta} \right) \frac{\dot{h}_m}{h_m}
  = - \frac{2 \gamma^2 \cos^2 \theta}{\alpha \rho g \cos\beta} \frac{\dot{h}_m}{h_m} ,
\end{equation}
where we have used $H_c=(2\gamma\cos\theta/\alpha \rho g)^{1/2}$.

The volume flux can be computed from Eqs. (\ref{eq:vol_cons}) and (\ref{eq:S_CPBS})
\begin{equation}
  \label{eq:Q_CPBS}
  Q = -\frac {\alpha H_c^4}{\cos^2\beta} \frac{\dot{h}_m}{h_m^2}
  \left(\ln \frac{h}{h_m} - \frac{h}{h_m} + 1 \right) .
\end{equation}
The dissipation function is given by
\begin{equation}
  \label{eq:Phi_CPBS}
  \Phi = \frac{1}{2} \int_{h_0}^{h_m} \ud h \frac{48\eta}{\alpha^3 S^4} Q^2 
  = \frac{24 B \eta}{\alpha} h_m \dot{h}_m^2 ,
\end{equation}
where $B$ is an integral
\begin{equation}
  B = \int_{\epsilon}^1 \frac{(\ln x - x +1)^2}{( 1- \frac{1}{x})^4} \ud x \simeq 0.0671.
\end{equation}
Here $\epsilon=h_0/h_m$ is a small number at late times, and the integral can be evaluated in the limit $\epsilon \rightarrow 0$.

Using Eq.~(\ref{eq:dotF_CPBS}) and Eq.~(\ref{eq:Phi_CPBS}), the Rayleighian is obtained
\begin{equation}
  \mathscr{R} = \dot{F} + \Phi =  
  - \frac{2 \gamma^2 \cos^2 \theta}{\alpha \rho g \cos\beta} \frac{\dot{h}_m}{h_m}
  + \frac{24 B \eta}{\alpha} h_m \dot{h}_m^2. 
\end{equation}
The time evolution of $h_m(t)$ is determined by the Onsager principle $\partial \mathscr{R}/\partial \dot{h}_m=0$
\begin{equation}
  \dot{h}_m = \frac{1}{24B} \frac{\gamma^2 \cos^2\theta}{\eta \rho g \cos\beta} \frac{1}{h_m^2} 
  \simeq 0.621 \frac {\gamma^2 \cos^2\theta }{ \eta \rho g \cos\beta} \frac {1}{h_m^2} .
\end{equation}
Taking the initial condition $h_{m}(t=0)=0$,  we obtain an analytical expressions
\begin{equation}
  \label{eq:hm_CPBS}
  h_m(t) = 1.23 \left( \frac {\gamma^2 \cos^2\theta}{\eta \rho g\cos\beta} \right)^{1/3} t^{1/3}.
\end{equation}

%

The result that $h_m(t)$ increases as $t^{1/3}$ is consistent with previous works \cite{Higuera2008, Ponomarenko2011}. 


\begin{figure}[htp]
  \includegraphics[width=0.8\textwidth]{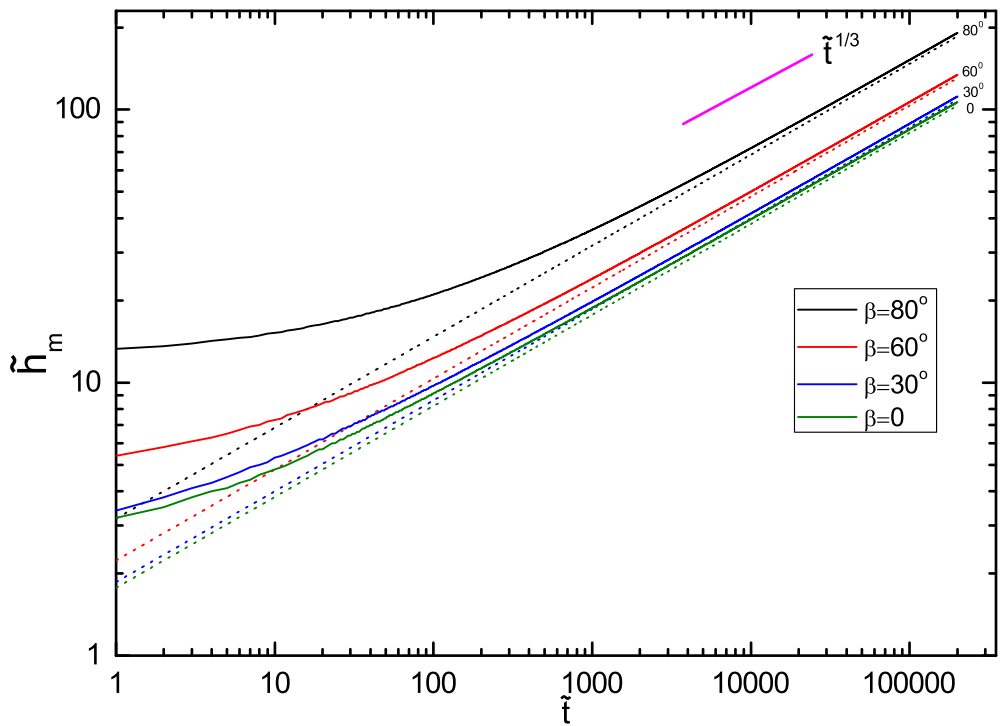}
  \caption{The length of meniscus front along the spine, $\tilde{h}_m$, is plotted against $\tilde{t}$  for different tilting angle $\beta$. The solid lines represent the numerical results based on  Eq.~(\ref{eq:evolution}), and the dot lines represent the analytical result (\ref{eq:hm_CPBS}).}
  \label{fig:3}
\end{figure}

The dimensionless version of Eq.~(\ref{eq:hm_CPBS}) takes the form   
\begin{equation}
  \label{eq:hm_CPBS2}
  \tilde{h}_m(\tilde{t})=\frac {1.77}{(\cos\beta)^{1/3}}\tilde{t}^{1/3} .
\end{equation}
We compare the numerical results of Eq. (\ref{eq:evolution}) and the analytical result (\ref{eq:hm_CPBS}) in Fig. \ref{fig:3}. 
The numerical results all approach the analytical result at later times for different value of $\beta$. 
For $\beta=0$, plates are vertically inserted into the liquid bath, and Eq.~(\ref{eq:hm_CPBS2}) becomes $\tilde{h}_m(\tilde{t})=1.77\, \tilde{t}^{1/3}$.
The result is very close to the self-similar solution $\tilde{h}_m(\tilde{t})=1.81\, \tilde{t}^{1/3}$ obtained by Higuera et al. \cite{Higuera2008}.

Equation (\ref{eq:hm_CPBS2}) indicates that the effect of tilting is to replace $g$ with $g \cos\beta$, which represents the gravity component along the spine.  
Although this result may look obvious, one must note that it is wrong to say that the effect of tilting is to replace $g$ by the effective gravity $g \cos\beta$.   
Indeed, the statement does not hold for the time evolution equation for $S(h,t)$.
Equation (\ref{eq:evolution0}) includes terms proportional to $g \sin\beta$.  
These terms arise from the fact that the center of mass of the liquid volume element at $h$ is different from the point at the spine (i.e., the center of mass is located at $(h, S(h)/2)$ and not at $(h,0)$).  
The deviation becomes larger as the tilting angle $\beta$ increases.  
On the other hand, the deviation becomes small for large values of $t$ since $S(h,t)$ becomes small (the meniscus becomes close to the spine) for large value of $t$.  
Hence the replacement of $g$ with $g \cos \beta$ is justified only in the asymptotic region of $t \gg t_c$.

\subsection{Model for Nepenthes alata}

In this section, we propose a simple model for the peristome surface of Nepenthes alata \cite{ChenHuawei2016}. 
To mimic the biological surface shown in Fig. {\ref{fig:4}(a), we consider a curved corner shown in Fig. {\ref{fig:4}(b) which captures the geometric essence of the peristome surface. 
The corner contains three parts: 
the first part is a straight line AB of length $L$ which is vertical to the liquid bath; 
the last part is also a straight line CD with a tilted angle $\varphi$;
in between AB and CD, a curved part BC has the shape of an arc, with the radius $R$ and the angle $\varphi$. 
Note for two flat plates, it is not possible to have a curved intersection. 
In our model, the two plates are formed by two lines that intersect on the curve ABCD with an angle $\alpha$ and perpendicular to the ABCD curve locally. 
For small $\alpha$, the two plates are almost flat. 

\begin{figure}[htp]
  \includegraphics[width=1.0\textwidth]{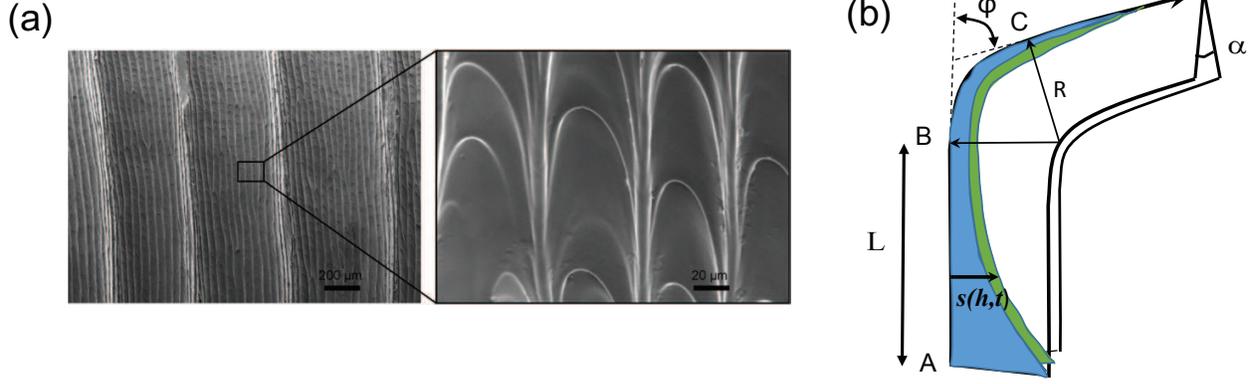}
  \caption{(a) Replicas of the peristome surface. (Reprinted from Ref.~\cite{ChenHuawei2016} with permission from the Springer Nature.) 
(b) Schematic picture of a bending model.}
  \label{fig:4}
\end{figure}

We use the curvilinear coordinates with $h$-axis following the corner line ABCD and $s$-axis perpendicular to $h$ locally.
The basic assumption is that the thickness of the liquid is much smaller than the curvature of the corner line, i.e., $S(h) \ll R$. 

The free energy of liquid in AB, BC and CD parts can be written as, respectively
\begin{eqnarray}
  F_{AB} &=& \int_{h_{0}}^{L} \ud h 
    \Bigg\{ \frac{1}{2}\rho g \alpha S^2 h-2\gamma S \cos\theta \Bigg\} \\
  F_{BC} &=&\int_{L}^{L+R\varphi} \ud h \Bigg\{ 
             \frac{1}{2} \rho g \alpha S^2 ( R \sin \frac{h-L}{R} + L ) 
             - \frac{1}{3} \rho g \alpha S^3 \sin \frac{h-L}{R} 
             - 2\gamma S \cos\theta \Bigg\} \\
  F_{CD} &=& \int_{L+R\varphi}^{h_m(t)} \ud h \Bigg\{  
      \frac{1}{2} \rho g \alpha S^2 \Big( L + R\sin\varphi + (h-L-R\varphi) 
             \cos\varphi \Big) \nonumber \\
             && - \frac{1}{3} \rho g \alpha S^3 \sin\varphi
             -2\gamma S \cos\theta \Bigg\} 
\end{eqnarray}
where $h_m(t)$ denotes the length of meniscus front along the spine for time $t$, 
$h_{0}$ is a certain reference point on AB selected at an early stage of rising.
This point is fixed as the initial boundary for the subsequent evolution equation. 

The equilibrium shapes of the meniscus are described by a set of dimensionless equations, which can be derived by the minimization of the total energy
\begin{eqnarray}
  \tilde{h}_{AB} &=& \frac {1}{\tilde{S}} , \\
  \tilde{h}_{BC} &=& \tilde{R} \sin^{-1} \Big[ \frac{1/\tilde{S}-\tilde{L}} {\tilde{R} - \tilde{S}} \Big] + \tilde{L} , \\
  \tilde{h}_{CD} &=& \frac {1}{\tilde{S}\cos \varphi} + \tilde{S} \tan \varphi 
                -\frac {\tilde{L}+\tilde{R}\sin\varphi} {\cos \varphi} 
                + \tilde{L} + \tilde{R}\varphi .
\end{eqnarray}
One can solve the above equations for $S$ as a function to $h$ to obtain the equilibrium profiles $S(h)$.

Now we derive the evolution equations. 
We can write the time derivative of the free energy using the volume conservation $\alpha S \dot{S} = -\partial Q / \partial h$ and then perform an integration-by-parts.
For the AB part 
\begin{equation}
  \dot{F}_{AB} = \int_{h_{0}}^{L} \ud h \Bigg\{ \rho g \alpha S h-2\gamma \cos\theta \Bigg\} \dot{S} = \int_{h_{0}}^{L} \ud h \Bigg\{ \rho g + \frac{2\gamma \cos\theta}{\alpha S^2} \frac{\partial S}{\partial h} \Bigg\} Q .
\end{equation}

For the BC part, we use a short-hand notation $\phi = \frac{h-L}{R}$,
\begin{eqnarray}
  \dot{F}_{BC} &=& \int_{L}^{L+R\varphi} \ud h \Bigg\{\rho g \alpha S 
                   (R \sin\phi + L) - \rho g \alpha S^2 \sin\phi 
                   - 2 \gamma \cos\theta \Bigg\} \dot{S} \nonumber \\
               &=& \int_{L}^{L+R\varphi} \ud h \Bigg\{\rho g \cos\phi 
                   - \rho g \sin\phi \frac{\partial S}{\partial h}
                   - \frac{\rho g}{R} \cos\phi S 
                   + \frac{2 \gamma \cos\theta}{\alpha S^2} \frac{\partial S}{\partial h}
                   \Bigg\} Q 
\end{eqnarray}

For the CD part
\begin{eqnarray}
  \dot{F}_{CD} &=& \int_{L+R\varphi}^{h_m(t)} \ud h \Bigg\{ \rho g \alpha S 
  (L + R\sin\varphi + (h-L-R\varphi) \cos\varphi) 
  - \rho g \alpha S^2 \sin\varphi - 2\gamma\cos\theta \Bigg\} \dot{S} \nonumber \\
  &=& \int_{L+R\varphi}^{h_m(t)}dh \Bigg\{ \rho g \cos \varphi  
  - \rho g \sin\varphi \frac{\partial S}{\partial h}
  + \frac{2 \gamma \cos\theta}{\alpha S} \frac{\partial S}{\partial h} \Bigg\} Q 
\end{eqnarray}

The dissipation function $\Phi$ has the same form as before 
\begin{equation}
  \Phi=\frac {1}{2}\int_{h_0}^{h_m(t)} \ud h \frac {48\eta}{\alpha^3 s^4}Q^2\, .    
\end{equation}

Onsager principle states that the time evolution of $Q$ is determined by $\delta (\dot{F}+\Phi)/\delta Q=0$. 
Using the expressions for $\dot{F}$ and $\Phi$, we can derive a set of equations that determine the time evolution of meniscus.
In the dimensionless form, these equations are
\begin{eqnarray}
  \label{eqT:ABe}
  \frac {\partial \tilde{S}}{\partial \tilde{t}} &=& 
    \tilde{S}^2 \frac{\partial \tilde{S}}{\partial \tilde{h}} 
    + \frac {1}{2} \bigg(\frac{\partial\tilde{S}}{\partial \tilde{h}}\bigg)^2 
    + \frac {1}{4} \tilde{S} \frac{\partial^2\tilde{S}}{\partial \tilde{h}^2} , \\
  \frac {\partial \tilde{S}}{\partial \tilde{t}} &=& 
     \frac {\tilde{S}^3}{4\tilde{R}} \sin\phi
     \bigg( \frac{\tilde{S}}{\tilde{R}}-1 \bigg) 
     + \tilde{S}^2 \cos\phi \bigg( 1 - \frac{3 \tilde{S}}{2 \tilde{R}} \bigg)
     \frac {\partial\tilde{s}}{\partial \tilde{h}} \nonumber \\ 
  \label{eqT:BCe}
  & & + \bigg( \frac{1}{2} - \tilde{S}^2 \sin\phi \bigg) 
      \bigg( \frac{\partial \tilde{S}}{\partial \tilde{h}} \bigg)^2 
      + \frac{1}{4} \tilde{S} \bigg( 1 - \tilde{S}^2 \sin\phi \bigg) 
      \frac {\partial^2\tilde{S}}{\partial \tilde{h}^2}  , \\
  \label{eqT:CDe}
  \frac {\partial \tilde{S}}{\partial \tilde{t}} &=& \tilde{S}^2 \cos\varphi
  \frac {\partial\tilde{s}}{\partial \tilde{h}} 
     + \bigg( \frac{1}{2} - \tilde{S}^2 \sin\varphi \bigg)
       \bigg( \frac{\partial \tilde{S}}{\partial \tilde{h}}\bigg)^2 
     + \frac{1}{4} \tilde{S} \bigg( 1 -  \tilde{S}^2 \sin\varphi \bigg)
       \frac{\partial^2 \tilde{S}}{\partial \tilde{h}^2}.
\end{eqnarray}
Equations (\ref{eqT:ABe}), (\ref{eqT:BCe}), and (\ref{eqT:CDe}) are the governing equations of meniscus in AB, BC, and CD part, respectively.

\begin{figure}[htp]
  \includegraphics[width=1.0\textwidth]{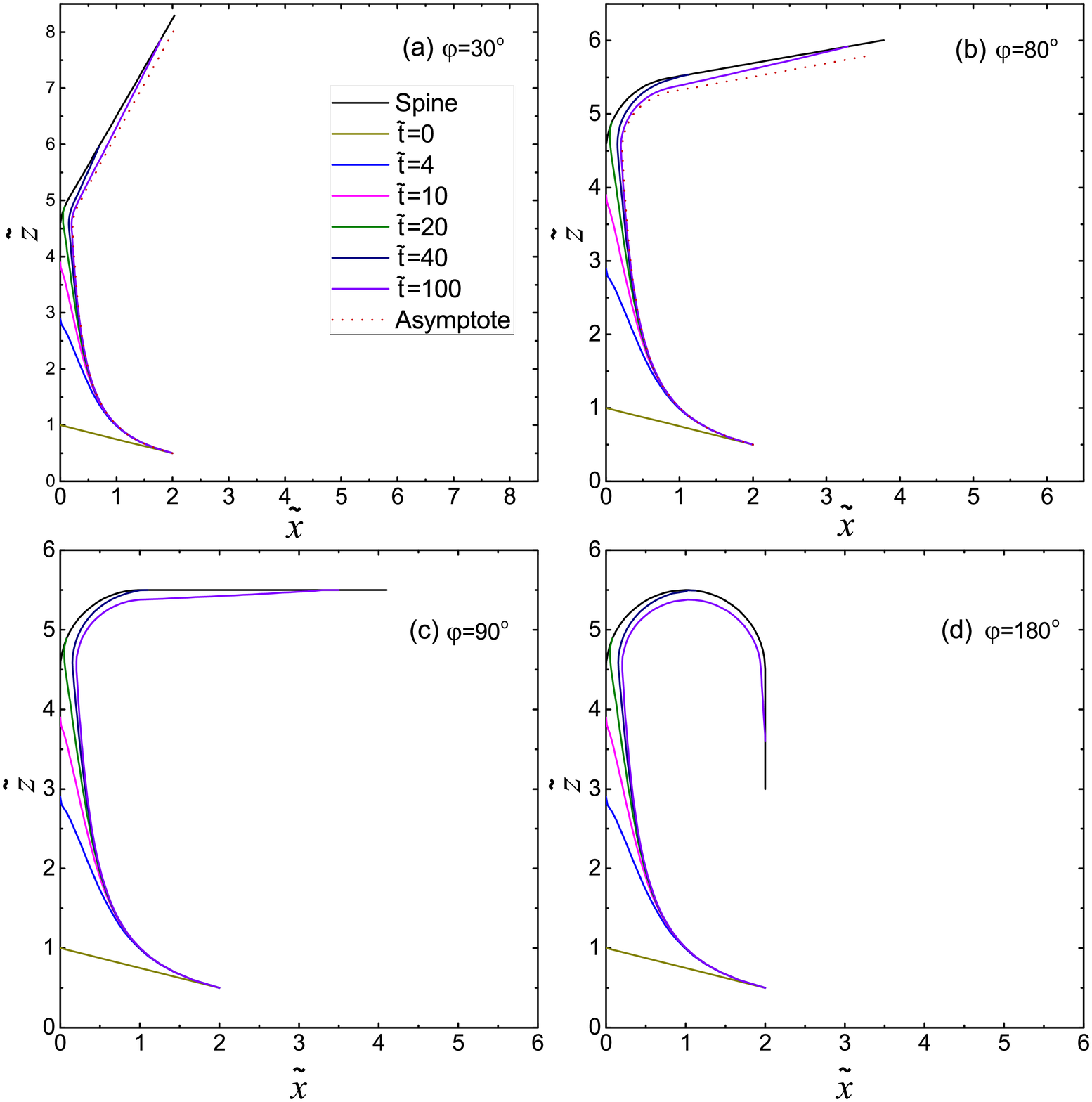}
  \caption{Profile of time evolution of meniscus for different bending angles: 
(a) $\varphi=30^{\circ}$, (b) $\varphi=80^{\circ}$, (c) $\varphi=90^{\circ}$, (d) $\varphi=180^{\circ}$. 
The dot lines denote corresponding equilibrium shape.}
  \label{fig:5}
\end{figure}

The time evolution of meniscus profiles for various bending angle $\varphi$ are shown in Fig.~\ref{fig:5} for $\tilde{L}=4.5$, $\tilde{R}=1$.
When $\varphi < 90^{\circ}$, meniscus climbs along the spine and tends to equilibrium as time increases.
When $\varphi > 90^{\circ}$, the equilibrium no longer exists and liquid tends to pour down along the corner at late times [Fig.~\ref{fig:5}(d)]. 
Gravity plays a significant role in this situation, and the supply of water will increases unboundedly with time, which promotes liquid transport efficiently.

\section{Summary}

We have studied the time evolution and equilibrium behaviors of meniscus for a wetting liquid at a narrow corner structured by two intersecting plates with varied tilting angles.
The main results from our numerical and analytical studies can be summarized as following:
\begin{itemize}
\item[(1)] For the equilibrium case, the meniscus shape is a hyperbolic function relevant to tilting angles.
\item[(2)] The length of meniscus front along the spine will increase as the cubic root of time and have a larger value for a smaller inclined angle predicted by an analytical expression. This is consistent with numerical \cite{Higuera2008} and experimental results \cite{Ponomarenko2011}.
\item[(3)] This work can be applied to a weakly bent corner and may be valuable for liquid transport in biology and medicine and engineering etc.
\end{itemize}

\begin{acknowledgement}
This work was supported by the National Natural Science Foundation of China (NSFC) through the Grant No. 21504004, 21574006, 21622401, and 21774004. 
M.D. acknowledges the financial support of the Chinese Central Government in the Thousand Talents Program.
\end{acknowledgement}


\section{Appendix: Effect of Initial Conditions }
\label{app:ic}
\begin{figure}[htp]
  \includegraphics[width=0.7\textwidth,draft=false]{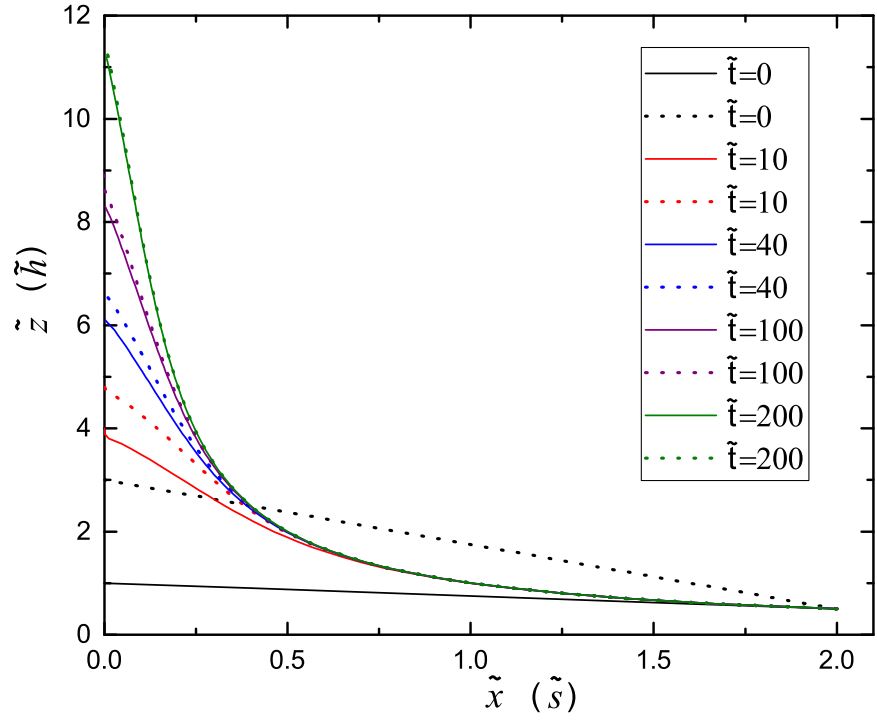}
  \caption{The time evolution of the meniscus for different initial conditions. The solid lines are the evolution results based on the initial condition (I)  $\tilde{s}=-4 \tilde{h}+4$,  and  the dot lines are the evolution results based on (II) $\tilde{s}=-0.8 \tilde{h}+2.4$.}
  \label{fig:6}
\end{figure}
Here we analyze the effect of initial conditions on the numerical results of Eq.~(\ref{eq:evolution}).
We use $\beta=0$ as an example. 
Two different initial profiles of the meniscus are considered for comparison, and both of them are straight lines: 
\begin{itemize}
\item[(I)]
\begin{equation}
  \tilde{s} = -4 \tilde{h}+4 ,
\end{equation}
\item[(II)] 
\begin{equation}
  \tilde{s} = -0.8 \tilde{h}+2.4.
\end{equation}
\end{itemize}

Figure \ref{fig:6} shows the time evolution of the profiles of the meniscus. 
One can see that after a relatively short time $\tilde{t}=40$, the profiles from two different initial conditions tend to nearly the same profile. 
Thus the initial conditions only affect the dynamics over a very short time scale, and have negligible influence on the long-time dynamics.


\bibliography{Taylor}




\end{document}